# Insights from Publishing Open Data in Industry-Academia Collaboration

Per Erik Strandberg[1], Philipp Peterseil[2], Julian Karoliny[3], Johanna Kallio[4], and Johannes Peltola[5]

## Abstract

Effective data management and sharing are critical success factors in industry-academia collaboration. This paper explores the motivations and lessons learned from publishing open data sets in such collaborations. Through a survey of participants in a European research project that published 13 data sets, and an analysis of metadata from almost 281 thousand datasets in Zenodo, we collected qualitative and quantitative results on motivations, achievements, research questions, licences and file types. Through inductive reasoning and statistical analysis we found that planning the data collection is essential, and that only few datasets (2.4%) had accompanying scripts for improved reuse. We also found that authors are not well aware of the importance of licences or which licence to choose. Finally, we found that data with a synthetic origin, collected with simulations and potentially mixed with real measurements, can be very meaningful, as predicted by Gartner and illustrated by many datasets collected in our research project.

## 1 Background

"Data is more valuable than oil," the Economist proclaimed last decade, following Humby who famously said that data is the new oil a decade earlier [1,6]. Data continues to be a critical success factor also in industry-academia collaboration (which refers to the partnership between industrial companies and academic institutions to leverage their respective strengths for mutual benefit) and research in general [4,9,10]. While the problem of data scarcity could possibly be reduced with data augmentation, e.g. by rotating images in a data set of images [2], this only limits the symptom and does not really solve the problem of lack of data in academia. Several previous studies have focused on lessons learned from industry-academia collaboration, e.g.: Marijan and Sen [10] that identified 14 patterns and 14 anti-patterns for industry-academia collaboration such as the positive effects of reciprocity and active dialogue, or Cederbladh et al. [4] who identified experiences and challenges when developing cyber-physical systems in industry collaborations such as the value of jointly formulating problems, or Garousi et al. [7] that identified that about half of the projects studied were initiated by industry. However, to the best of our knowledge, no previous work has focused on aspects of publishing open data sets from industry-academia collaborations. Therefore, in this paper we explore the motivations and lessons learned when partners published data sets. By understanding these motivations and lessons, we aim to enhance data collection practices, improve the reuse of datasets,

[1] Corresponding author: per.strandberg@westermo.com. Westermo Network Technologies AB, 721 30 Västerås, Sweden. ORCID: 0000-0003-1688-6937
[2] Johannes Kepler University Linz, Altenberger Straße 69, 4040 Linz, Austria. ORCID: 0000-0003-2536-0515
[3] Silicon Austria Labs GmbH, Altenberger Straße 66c, 4040 Linz, Austria. ORCID: 0000-0001-8150-382X
[4] VTT Technical Research Centre of Finland Ltd., Kaitoväylä 1, P.O. Box 1100, 90571 Oulu, Finland. ORCID: 0000-0002-7408-170X
[5] VTT Technical Research Centre of Finland Ltd., Kaitoväylä 1, P.O. Box 1100, 90571 Oulu, Finland. ORCID: 0000-0002-6248-152X

and promote the effective use of synthetic data. We did this by asking partners of the European collaborative industry-academia research project InSecTT[6] in a survey, and to enhance these findings we also analysed metadata from more than 280 732 data sets from Zenodo.

The main contributions of this paper are:
1. Qualitative and quantitative insights of the motivations and methods for publishing datasets in an industry-academia research project: e.g. that planning the data collection process is often more challenging than the technical aspects, highlighting the importance of workflow management and stakeholder involvement.
2. Quantitative insights from hundreds of thousands of Zenodo data sets: e.g. that less than 2.4% of Zenodo datasets are accompanied by scripts that improve reuse, indicating a significant gap in the provision of supporting tools and software.
3. Recommendations for publications of future data sets: e.g. we recommend using both GitHub and Zenodo for publishing data to leverage GitHub's integration capabilities and Zenodo's DOI assignment, ensuring both accessibility and permanence.

The remainder of this paper is organised as follows. Section 2 presents the survey we used, the identified datasets from the InSecTT project, as well as how we scraped Zenodo for metadata. Section 3 presents the results from the survey and analysis of metadata. In Section 4 we discuss the results and Section 5 concludes the paper. For reference, the survey questions are presented in Appendix A.

## 2 Materials and Methods

**Table 1**. Overview of InSecTT datasets.

| Dataset | Context | Accessibility |
| --- | --- | --- |
| UWB[7] weak-NLOS[8] structured dataset | Static UWB measurements in different weak-NLOS environments | https://github.com/ppeterseil/UWB-weak-NLOS-structured-dataset |
| UWB dynamic localization dataset | Dynamic UWB measurements evaluating autoencoder-based trustworthiness indicator | https://github.com/ppeterseil/UWB-AEC-trust-dataset |
| AD-EYE[9] open-KTH[10] | LIDAR[11] and other vehicular sensor data | https://www.adeye.se/open-kth |
| Modular Ice-cream factory dataset on anomalies in sensors | Sensor data with and without anomalies from a factory | https://github.com/vujicictijana/MIDAS |
| ICSFlow[12]: an open-source | Network data and process state | https://www.kaggle.com/datasets/ali |

---

[6] InSecTT: Intelligent Secure Trustable Things. InSecTT was a three-year project funded by the ECSEL Joint Undertaking (JU) and EU Horizon 2020. https://www.insectt.eu/

[7] UWB: Ultra-wideband wireless technology
[8] NLOS: Non-line-of-sight
[9] AD-EYE: Testbed for Automated Driving and Intelligent Transportation Systems
[10] KTH: Royal Institute of Technology (Swedish: Kungliga Tekniska Högskolan)
[11] LIDAR: Light (or laser) detection and ranging
[12] ICS: Industrial Communication System

| dataset for intrusion detection purposes | variables logs | rezadehlaghi/icssim |
|---|---|---|
| 5G Measurement data | 5G channel parameters and uplink throughput. | https://github.com/MTU-Insectt/Measurements5G |
| BLE[13] Indoor zone-based localization dataset | Indoor localization with BLE Received Signal Strength Indicator | https://doi.org/10.5281/zenodo.4073072 |
| WSN[14] Power consumption dataset | Wireless sensor network power consumption for four protocols | https://doi.org/10.5281/zenodo.7762712 |
| TDMA[15] interference dataset | Periodic interference measurements with a TDMA network | https://doi.org/10.5281/zenodo.6306288 |
| BLE Channel Sniff Dataset | BLE traffic measurements | https://doi.org/10.5281/zenodo.7152044 |
| The Westermo network traffic dataset | Network data with anomalies | https://github.com/westermo/network-traffic-dataset |
| Air quality data set | Airport indoor air quality metrics | https://iotlab.unipr.it/ https://data.mendeley.com/datasets/bv2hvm4pmz |
| MOTSynth[16] dataset | Computer vision pedestrian tracking | https://motchallenge.net/data/MOTSynth-MOT-CVPR22/ |

The empirical data about the InSecTT partners' experiences and learnings in collecting data sets were gathered via a survey. We designed the survey questions related to datasets, including (1) their publication process, (2) purpose of the dataset, (3) motivation of collecting the dataset, (4) research questions or other targets studied with the dataset, (5) type and format of data, (6) how the dataset was collected, (7) whether supporting software or tools were published, (8) the main achievements and results from the dataset, and (9) learnings from the data collection process (see Appendix A for the survey questions). All these questions were open-ended. In addition, demographics, including type of organization and working experience in years were asked. After drafting the questionnaire, two senior researchers with over 20 years of experience from participating organizations reviewed it and revised it to be more explicit.

The survey data was collected anonymously in winter 2023 - 2024 using the Questback Inc. online survey tool. The survey was distributed via the InSecTT mailing list and obtained a total of eleven responses (some covering multiple data sets), two of which are on non-published datasets (excluded in Table 1), for a total of thirteen open datasets. We used inductive reasoning to analyze written survey responses. The primary objective was to observe experiences and actions, address challenges and derive lessons for future data collection activities. Inductive reasoning allowed us to identify prominent, frequent, or significant themes from the data without constraints from structured methods [15]. Table 1 describes the thirteen datasets of the InSecTT project with name, context and link for download. The datasets include UWB measurements, vehicular sensor data, factory sensor anomalies, network traffic data, air quality metrics and video for pedestrian tracking.

---

[13] BLE: Bluetooth Low Energy
[14] WSN: Wireless sensor network
[15] TDMA: Time Division Multiple Access
[16] MOT: Multiple Object Tracking

Although the number of datasets within the InSecTT project was quite significant, it remains somewhat limited for making statistical assertions. For instance, our survey enabled us to comment on the distribution of used file formats, however, we could not correlate these findings with the general practices within the dataset community. Therefore, we extended the evaluation of this work with statistical data from Zenodo, a widely used online platform for hosting datasets and other types of research output. Zenodo itself does not directly offer statistics about their published datasets, but they provide an API to access metadata of individual datasets. Consequently, we developed a Python script to collect the metadata like publication date, used licence, and file extension of all dataset entries in Zenodo between 2000 and 2023. This allows us to relate and categorise the results of our survey questions. Figure 1 shows the distribution of the 280,732 datasets gathered from Zenodo depicted over the publication year. This already highlights the increasing trend of making data publicly available as an initial result.

Secondary data from both the survey and the Zenodo analysis has been published at Zenodo [14].

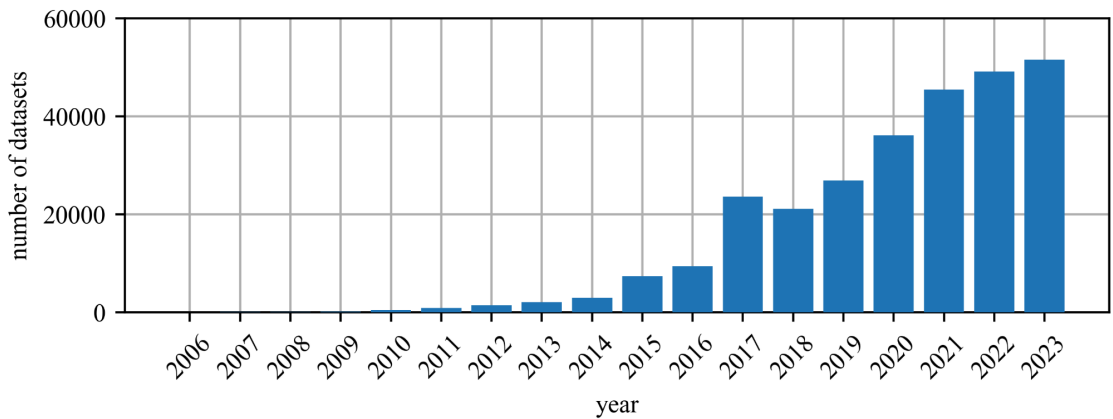

**Figure 1**: New Zenodo data sets per year.

## 3 Results

In this section we present the results from the InSecTT survey and the Zenodo analysis: demography, the data collection processes used by the respondents, where data was published, the licences used, how supporting tools or software was used, the original purpose and motivation for publishing the data, research questions they aim to answer, achievements and lessons learned.

### 3.1 Demography

Participants (n=11) were voluntarily recruited from InSecTT project partners by sending an invitation via email to the mailing list of the InSecTT project in November 2023. Among the survey respondents, 54.5% worked in universities, 27.3% in large enterprises (defined by the European Commission as companies with at least 250 employees), and 18.2% in research institutes. Nearly half of the respondents (45.5%) had 16-20 years of working experience, while the proportions for 1-5 years, 6-10 years, and 11-15 years of working experience were all approximately one fifth of the respondents (18.2%).

## 3.2 Data Collection Processes

We asked the respondents how the datasets had been collected, and asked them to describe the environment and equipment used, as well as the use of any tools, software or data storage approaches. Most (8 of 11 respondents) used real hardware during the collection, e.g. vehicles, sensors, or network equipment. Three data sets instead used only simulators, ICS-Flow simulated network nodes, MOTSynth used the Rockstar Advanced Game Engine (RAGE) known from Grand Theft Auto V to simulate people moving, and MIDAS simulated sensors in an ice cream factory. Some (e.g. the Westermo network traffic dataset) connected simulators to physical devices. An advantage of a simulated environment is that anomalies and attacks can be injected and labelling is simplified. In order to collect diverse data, the respondents mentioned that experiments were often repeated with slight variations, e.g. by changing the clothes of people and the weather in RAGE, or changing the environment when collecting sensor data (indoors, outdoors or anechoic chamber). Three respondents mentioned cleaning data: one data set was cleaned to avoid company data to leak, another cleaned to ensure no controversial images were kept (e.g. vulgar graffiti in RAGE), and the final one cleaned lidar data of vulnerable or identifiable participants.

## 3.3 Publication: Location and Data Formats

A detailed examination of the datasets linked to the InSecTT project indicates that a significant portion, specifically 38.5% (5 out of 13 datasets), was made available on Zenodo. Additionally, a smaller fraction, 7.7% (1 out of 13), found their way onto Mendeley Data and Kaggle, each. It's particularly interesting to note that 38.5% (5 out of 13) of the datasets were hosted on GitHub. While GitHub is not considered a suitable platform for dataset publication, this trend suggests that research groups appreciate GitHub for staging datasets in preparation for their actual publication. Transferring to an open access data repository, e.g. Zenodo, at a later stage is beneficial, as they typically not only archive the work status but also assign a DOI to the content. This process enhances the publication's traceability and ensures its preservation, facilitating easy access and citation by the academic community. Table 2 gives an overview about most prominent platforms that are not limited to a specific field of research.

**Table 2**. Overview of dataset publication locations based on Stall et al. [13].

| Platform | Size limit | Note | Fee | Identifier | Usage in InSecTT |
|---|---|---|---|---|---|
| Zenodo | 50 GB | - | Free | DOI | 38.5% |
| Figshare | unlimited (public sets) | 20 GB/file up to 5 TB/file (with fee) | Depends on file size | DOI | - |
| Dryad | 300 GB | CC0 exclusively | $150 | DOI | - |
| Mendeley Data | 10 GB | - | Free | DOI | 7.7% |
| OSF | Unlimited | 5 GB/file | Free | DOI | - |
| Harvard Dataverse | Unlimited | 2.5 GB/file | Free | DOI | - |
| Kaggle | 100 GB | Associated with data science competition | Free | DOI | 7.7% |
| GitHub | 5 GB | recommended less than 50 MB/file, less than 1 GB total | Free | - | 38.5% |
| Custom Project Homepage | - | - | - | - | 7.7% |

The most common file format in InSecTT was comma separated values (CSV) with seven of eleven data sets using it. The respondents mentioned that this was a data format that could easily be opened and parsed by others. Two datasets involved captured network packets in a raw format (PCAP), and both these were accompanied with processed network traffic in flows [3] in CSV format. Other formats include raw and annotated video, Python pickle, Robot Operating System bag (rosbag), Postgres database, and Point Cloud Data (pdc).

An interesting observation is that some data sets relied on binary formats of specific versions of applications (e.g. a certain Python version). Furthermore, when collecting network traffic, privacy could be a problem, in particular if the recording is done on a regular laptop or if wireless traffic is captured - unintended traffic might have found its way into the dataset. We also observe that large data sets (e.g. in CSV format) may not be published at GitHub since there are limitations on sizes of individual files.

Through an analysis of the data from Zenodo we make complementary observations. The majority of published files were in the form of archives, the contents of which remain undisclosed to us. Following these, the most prevalent file formats were structured text formats, such as CSV, JSON, and XML, succeeded by various binary file formats, including SAV, DAT, H5, and MAT. Notably, about a quarter of the data sets (23.6%) had at least one text document (e.g., MD, DOC, DOCX), which we believe serve as descriptions for the datasets. Additionally, there is a significant presence of image files, which are presumably either part of machine learning datasets or pictures used in the dataset descriptions. See overview in Figure 2.

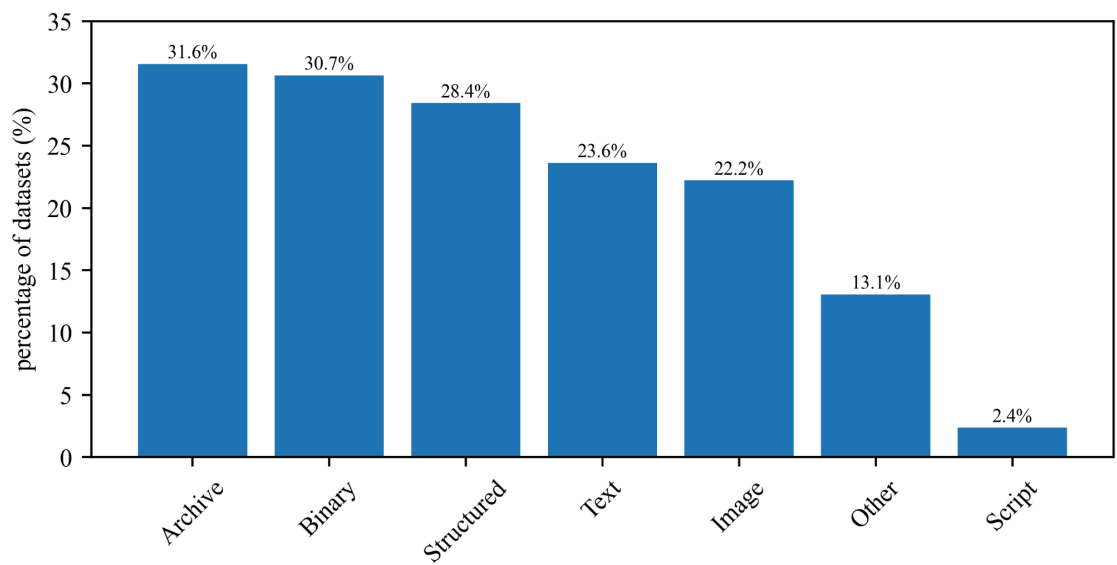

**Figure 2**: Distribution of file formats used in Zenodo data sets.
Values show the percentages of data sets that have at least one file of this type.

## 3.4 Licences

Licences in open-source datasets are very important as they define the terms under which the data can be used, shared, and modified and promote collaboration and transparency. Specific types of open source licences are the copyleft licences, such as the GNU General Public License (GPL). They ensure that any derivative works are also distributed under the same or compatible licence terms and thereby preserve openness. While the copyleft licences are often favoured for publishing open software, they can be too restrictive for open datasets. For datasets often licences with fewer restrictions such as the Creative Commons Attribution (CC BY) licence is favourable, as it requires only attribution to the original creator. Examining the datasets of the InSecTT project reveals that the majority of them were published under CC-BY (7 of 13 datasets, or 54%), highlighting the recognition of the advantages associated with its use for datasets. Four datasets (4 of 13 datasets, or 31%) were published without any licence. Here the use without any agreement with the publisher is theoretically not possible, contradicting the idea of open datasets. Two partners (2 of 13 datasets, or 15%) used other licences like Eclipse Public License (EPL) and GPL, which are typically used for software licensing. While both licences promote open-source collaboration, their copyleft requirements—strict for GPL and more permissive for EPL— can complicate the use of open data sets by imposing restrictions that might not align with the intention of providing open-source datasets.

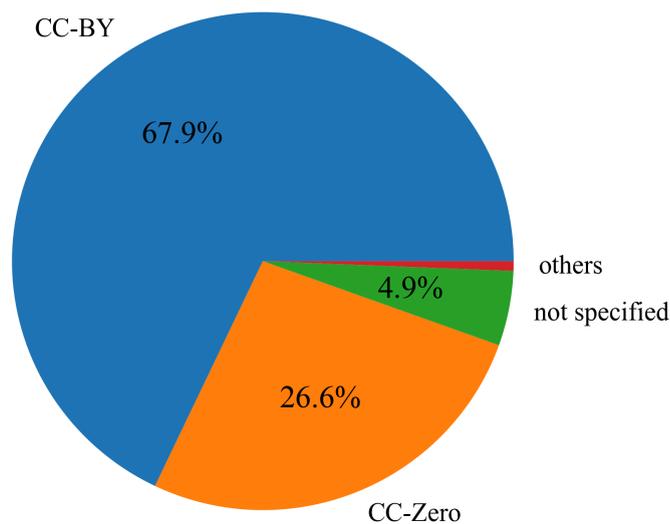

**Figure 3**: Distribution of licences used in Zenodo data sets.

The topic of licences was a significant motivator for evaluating the published datasets on Zenodo, with the results highlighted in Figure 3. The figure shows the distribution of the used licences from the metadata of the datasets. Also here the majority (67.9%) of all published datasets utilised the CC-BY licence, where we also included all additions like non-commercial, share-alike, or no-derivatives. When adding a new dataset to Zenodo, CC-BY is the default choice and requires active modification to change. In comparison to InSecTT, only 4.9% of the datasets lack a clearly defined licence, which is significantly lower. One reason could be Zenodo's default behaviour when adding a dataset, but it might also reflect a lack of knowledge about datasets and licences among InSecTT partners. Only 0.6% of the datasets used other licences, representing a group where the individual licences would not constitute a representative percentage. This includes for example MIT,

Apache, BSD, but also EPL and GPL as in the InSecTT datasets. In contrast to Zenodo where 26.6% of the datasets used CC-Zero, no InSecTT dataset was published under this licence. Although the number of InSecTT datasets is not representative, one reason could be the strong research orientation of most InSecTT partners. CC-Zero permits the use of the dataset without attributing the original author, which may be unfavourable in the research community.

### 3.5 Supporting Tools and Software

Most data sets (7 responses or 63.6%) were published without supporting software or tools. Some data (3 responses or 27.3%) added code for parsing the data or other example code. Two (18.2%) published the data with algorithms that repeated some experiments, and one additional respondent mentioned that it would have been best practice to do so, but was unable due to time constraints. When compared with the results from Zenodo, we can only establish an upper bound for supporting software. As illustrated in Figure 2, approximately 2.4% of the datasets in Zenodo contain files that we identify as scripts (e.g., .py, .m, or .R). Although supporting software can improve the usability of open datasets, our findings indicate that its inclusion is not a widespread practice. Moreover, the actual number of supporting software could be even lower, as we have no means to verify whether all identified scripts truly function as supporting software.

### 3.6 Original Purpose and Research Questions

The main purposes for collecting datasets during the project were (1) to support ongoing research, (2) to facilitate AI tool development within either technical or use case-specific perspectives, and (3) to ensure research reproducibility. As can be expected based on the scope of the InSecTT project, ML and related fields dominate.

We collected information about the research questions for classifying them and evaluating if the type of research questions has strong similarities across different published datasets in our survey. We used two different methods for classifying the questions. First we used a scheme by Shaw [12] for analysing the category of each identified research question. To get a more comprehensive view of the similarities, we also used a second classification scheme by Dillon [5].

Evaluating research questions is important as they provide information about the goals, motivations, approach, and scope of research for associated datasets. The classification can indicate if publishing open datasets is typical for a specific type of research domain and what are the typical research questions in the context of published datasets.

The classification by Shaw uses five main categories for classifying research questions. The first category refers to questions related to methods or means for developing new solutions or improvements over existing ones. The second category contains methods for analysing existing solutions. The third focuses on the design and evaluation of a particular instance, while the fourth describes the generalisation or characterization. Finally, the fifth category is about feasibility studies or exploration.

We identified that all categories defined by Shaw are present in the published datasets. Most typical types of questions were the third (7 RQs) and first (5 RQs) categories. The rest had either two or three research questions in each category.

The model that Dillon is using has four main categories for the research questions. (1) Descriptive questions that seek to identify the characteristics of a phenomenon or population. (2) Comparative questions that aim to compare and contrast two or more groups or conditions. (3) Relational questions that investigate the relationships between variables. (4) Causal questions that focus on understanding the cause-and-effect relationships between variables.

We found a spread of research questions: eight on first order or properties (e.g., what are the features needed to train a trustworthy ML solution in a certain wireless domain), five on the second order or comparisons (e.g., is method A better than method B for AI-powered anomaly detection in a certain domain), and seven were on third order RQs or relations (e.g. can we predict data rates given measurements on lost connection).

The results of both classification schemes are summarised in the following table that shows three dominating cross sections with more that 3 RQ's combining both schemes.

Table 3. Classifications of research questions posed to be answered by InSecTT datasets.

|  | Dillon (1984) [5] | | | |
| --- | --- | --- | --- | --- |
| **Shaw (2003) [12]** | **Properties** | **Comparisons** | **Conditionality** | **Sum** |
| Method or means of development | 4 | - | 1 | 5 |
| Method for analysis or evaluation | **2** | 1 | - | 3 |
| Design, evaluation, or analysis of a particular instance | - | **4** | **3** | 7 |
| Generalization or characterization | 1 | - | 2 | 3 |
| Feasibility study or exploration | 1 | - | 1 | 2 |
| Sum | 8 | 5 | 7 | |

Three examples from the dominating cells are
1. Method or means of development, while focusing on properties: provide an extensive collection of datasets for future developments
2. Design, evaluation, or analysis of a particular instance, while focusing on comparisons: How different supervised machine learning algorithms are performing for anomaly detection and classification in analog sensor data, considering classification performance and time consumption?
3. Design, evaluation, or analysis of a particular instance, while focusing on conditionality: how could you combine a rule-based network anomaly detection with a learning system?

### 3.7 Achievements

When we asked the respondents about the main achievements that came from the publication of the dataset, the most common reply is academic peer reviewed publications. Many have published the dataset along with academic publications, whereas a few published an article with the dataset as the core. Some academic publications also receive many citations and MOTSynth is an example of this. The dataset has already been cited over 100 times.

Academic publications are a consequence of research goals being achieved, and this is the second most common achievement mentioned by respondents. Examples include gaining new knowledge and the development of new methods.

Two respondents explicitly mention that publishing public datasets can improve the quality of research. One respondent mentions that relevant datasets from industry could lead to academics focusing on more relevant research (i.e. "closer to reality"), and another mentions transparent research and thus implicitly also reproducibility.

Finally, datasets have also led to commercializations, e.g. the MOTSynth data set has been used in many camera surveillance applications for pretraining.

### 3.8 Lessons Learned

Four respondents specifically mentioned that planning the data collection was essential, one wrote "workflow issues were bigger than the technical challenges, which were easily overcomed" and another mentioned that they spent a lot of effort restructuring the data to make it meaningful when outside of the original research context. When planning, researchers should consider the entire workflow, and design sensitive things out of the data (e.g., things that might otherwise require ethical approvals). If involving a company then one needs also to think about a process for information security, and involve the right stakeholders.

A few respondents mentioned cleaning data, this could be related to vulnerable individuals being captured in the data, protected places, or offensive contents. In terms of documentation, it was recommended to document as soon as possible, and that a readme-file in markdown format was a reasonable starting point.

Finally, the respondents mentioned that data is indeed useful for academia, for industry-academia collaborations, and that reviewers appreciate publications with open datasets.

## 4 Discussion

### 4.1 Research Questions

We found that the dominating RQs in the InSecTT project investigated methods or means of development, or design evaluation or analysis of a particular instance while focusing on properties, comparisons or conditionality. In other words, the main motivation for data collection and publication was to create data that allow developing new technical solutions or create improved functionality for existing solutions. Datasets allow understanding the properties of the phenomena or provide information on how to compare and predict technological performance.

The least popular RQs were on generalization, characterization, feasibility or exploration. We speculate that the reason these topics were less investigated is the strong focus on applied science in an industry-academia collaboration such as InSecTT. Furthermore, in some cases the research question was not exactly a real question which could indicate that the creation of the dataset has a clear technical requirement for developing a specific functionality. We would welcome clearer RQs and stronger links between RQ and data.

## 4.2 Synthetic Data?

Data availability is a challenge in many domains, and generating synthetic data is a tempting approach. According to the American technological research and consulting company Gartner [8], there is a shift towards using synthetic data for AI, from 1% in 2021 to 60% 2024. Among the InSecTT datasets, MOTSynth simulates real people to train computer vision AI in various scenarios. The simulation ensures that there are no security, GDPR or other privacy violations, and that ground truth in the form of labels is generated with the dataset. If humans would annotate the data, then labelling could be flawed. For a more general perspective of data, and anomaly detection, Schmidl et al. and Wenig et al. [11, 16] made a large scale analysis of 976 time series data sets from a wide range of domains. They used or re-implemented 71 anomaly detection methods and found that AI-based approaches are hardly better than simple methods. One challenge is the lack of labelled data sets, and for this reason they implemented the TimeEval toolkit for generating labelled time series with anomaly injection as well as tools for anomaly detection.

For the three InSecTT datasets MIDAS, ICSFLOW and Westermo network traffic, a factory is simulated and sensor data or network traffic is collected. In addition, the factory or the network is disturbed with anomalies. In a similar way this leads to trustworthy labels in the data while not disturbing an actual factory. We speculate that this trend of combining simulations, anomalies and partial use of hardware lowers the bar for data collection and publishing among industry.

Bansal et al. [2], conducted a systematic literature review on how to combat data scarcity. They propose to conduct (1) cost sensitive learning, an optimization strategy for better training data with class imbalance; (2) transfer knowledge, i.e., to transfer knowledge from models or domain experts, and also (3) data augmentation, i.e., generating new data based on existing data, e.g., by rotating images. We speculate that data augmentation could be a way to generate improved data sets: based on existing data from the real world, new data points from corner cases, adversarial examples or simply variants of existing cases could be generated.

## 4.3 Limitations, Validity Discussion and Future Work

While this study provided insights and practical recommendations for sharing open data sets in industry-academy collaboration, it is noteworthy to mention some limitations which may affect the validity of the study. The conducted survey was qualitative, which is vulnerable to subjective bias in the planning and analysis phases. Moreover, the sample of survey respondents was relatively small and restricted to one project, limiting the generalizability of the survey results. However, despite the limited number of respondents they cover all datasets of the InSecTT project. Additionally, the survey responses may be subject to biases, such as too optimistic or socially desirable answers. Nonetheless, this study provides valuable information to consider when planning and publishing open data sets.

This study focuses on datasets published in InSecTT and on Zenodo, which may not reflect the full range of data sets published across different platforms (see Table 2). Additionally, the findings are based on data collected up to 2023, and may become outdated as data publication practices and technologies evolve. We did not explore the contents of the datasets in detail, relying instead on metadata analysis. Future research could explore a larger number of platforms, examine dataset contents more thoroughly, or repeat the study at a later stage. Expanding the scope to include multiple projects could provide a more comprehensive understanding of data publication practices across different contexts. Conducting content analysis of the datasets, examining not only the metadata but also the quality and usability of the data could provide insights into what factors enhance dataset utility. Additionally, investigating the impact of quality aspects (e.g. inclusion of example code) could explain how quality enhance the reuse and citation of datasets. E.g., datasets accompanied by scripts or comprehensive documentation might be used more frequently.

# 5 Conclusions

In this paper, we present insights from publishing open data in industry-academia collaboration gathered through a survey with participants in the InSecTT project. The analysis shows that planning data collection in InSecTT could be a greater problem than technical issues. For comparison, we also mined metadata from almost 281 thousand datasets in Zenodo and found that very few Zenodo datasets (less than 2.4%) were accompanied with scripts that would improve reuse. Furthermore, we found that many InSecTT datasets were published without a licence and that most Zenodo datasets used the default choice (CC-BY). Finally, we found that publishing synthetic data can be very meaningful, as predicted by Gartner and illustrated with the MOTSynth dataset for pedestrian tracking in computer vision.

## 5.1 Recommendations

For publishing future data sets in industry-academia collaborations we recommend to:
- Create a research question and analyse it using a common taxonomy provided by e.g. Dillon or Shaw [5,12]. Verify that the result of the analysis matches the main goal of the activity.
- Plan the entire workflow before collecting data; design for easy cleaning or avoid collecting things that need to be filtered out.
- Document the data early, a readme in markdown is a suitable lowest effort format.
- If applicable, choose a non proprietary file format that is well-known and simple in favour of file size.
- Include source code illustrating how to parse and preferably conduct at least some experiment with the data. Ideally include a complete replication package. Especially synthetic datasets can benefit from supporting software.
- Consider synthetic data, or augmenting the data with semi-synthetic scenarios.
- Use both GitHub and Zenodo to publish data. GitHub facilitates easy integration of the dataset into other work and fosters possibilities for future collaborations. Zenodo assigns a DOI to the content enabling citing in publications and offers a direct Github interaction - datasets published in Github can be easily imported into Zenodo. Using only Github has the significant disadvantage that the content is not permanent and can be modified afterwards. For datasets, this is not beneficial and Zenodos offers a guaranteed version of the data with its DOI.
- Regarding licences, we advise against using copyleft (e.g. GPL) ones, as they are too restrictive for datasets. Licences like CC-BY, are ideal for datasets because they only require author attribution, aligning well with the DOI generated by Zenodo.

# Ethics Statement

The current work does not involve animal experiments, or any data collected from social media platforms. Survey participants contributed after informed consent regarding scope and purpose and no personal identifiers or data was collected.

# CRediT Author Statement

**All**: Investigation, Writing - Original Draft and Writing - Review & Editing. **PES, JoK and JP**: Conceptualization and Methodology. **PP and JuK**: Software and Visualization.

# Acknowledgements

The authors would like to thank partners from the InSecTT project for their contributions in answering the survey. The project leading to this application has received funding from the ECSEL Joint Undertaking (JU) under grant agreement No 876038 (InSecTT). The JU receives support from the European Union's Horizon 2020 research and innovation programme and Austria, Sweden, Spain, Italy, France, Portugal, Ireland, Finland, Slovenia, Poland, Netherlands, Turkey. This project has received funding from the ECSEL Joint Undertaking (JU) under grant agreement No 101007350 (AIDOaRt). The JU receives support from the European Union's Horizon 2020 research and innovation programme and Sweden, Austria, Czech Republic, Finland, France, Italy, Spain. This research was partially funded by HIPE project, which receives funding from VTT Technical Research Centre of Finland Ltd. and Business Finland.

# Declaration of Competing Interests

The authors declare that they have no known competing financial interests or personal relationships that could have appeared to influence the work reported in this paper.

# Appendix A: Survey Questions

1. Which type of organization are you working in? (Small or medium enterprise, Large enterprise, University, Research institute, Public sector or Other)
2. How long is your working experience? (1-5 years, 6-10 years, 11-15 years, 16-20 years, 21-25 years, 26-30 years, or Over 30 years)
3. Which InSecTT dataset did you contribute to?
4. Where has the data been published, and what was the process for publishing it?
5. What has been the original purpose of the dataset?
6. What has been the main motivation to publish the dataset?
7. Which have been the research questions or other challenge targets studied with the dataset?
8. What was the type and format of data?
9. How has the dataset been collected? Please briefly describe the environment (lab vs real-life), utilised equipment, software/tools and data storage approach(es), and data acquisition duration.
10. Have you published supporting software or tools with your dataset? If yes, please describe in short.
11. Which are the main achievements and results sprung up from the dataset so far? These include, but are not limited to, publications, commercialization, and exploitation plans.
12. Have you learnt some specific lessons during the data collection process? For example, share your experiences (in brief) about encouraging the stakeholder(s) to get involved , ways/workflows in solving encountered technical challenges, discovered best practices on data management/cleaning/processing/anonymization, viable solutions applied on ethical challenges (if any), the practical nuts and bolts of composing documentation/guidelines, and so on.